\documentclass[letterpaper,notoc]{CECS}

\title{de Sitter black hole with a conformally coupled scalar field in four
dimensions}
\author{Cristi\'{a}n Mart\'{\i }nez, Ricardo Troncoso and Jorge Zanelli\footnote{{\it E-mail:} {\tt  martinez@cecs.cl, ratron@cecs.cl, jz@cecs.cl}}\\ Centro de Estudios Cient\'{\i}ficos (CECS), Casilla 1469, Valdivia, Chile.  }

\preprint{{\tiny CECS-PHY-02/05} }

\abstract{
A four-dimensional black hole solution of the Einstein equations with a
positive cosmological constant coupled to a conformal scalar field is given.
There is a curvature singularity at the origin, and the scalar field has a
divergence inside the event horizon. The electrically charged solution,
which has a fixed charge-to-mass ratio, is also found. The quartic
self-interacting coupling constant becomes bounded in terms of Newton's
constant and the cosmological constant. The solution satisfies both dominant
and the strong energy conditions.}

\begin{document}

In this paper we report an exact black hole solution in four dimensions with
electromagnetic and conformally coupled scalar fields. This black hole only
exists for a positive cosmological constant $\Lambda $, in agreement with
recent observations \cite{PositiveLambda}, and if a quartic self-interaction
coupling is considered. This coupling does not spoil the conformal
invariance of the scalar field equation. There is a curvature singularity at
the origin, and the scalar field is regular both on and outside the event
horizon. The electrically charged solution exists with a fixed
charge-to-mass ratio, which is given in terms of the fundamental constants
of the action. This last feature implies a bound for the self-coupling
constant, which is also obtained through the analysis of the solution with a
nonvanishing constant scalar field.

Static scalar field configurations such as those presented here, which are
regular both at the horizon as well as outisde, are unexpected in view of
the no-hair conjecture (see, e.g., \cite{Heusler}). Furthermore, this
solution can be seen to satisfy the dominant and the strong energy
conditions, which corresponds to real matter under normal conditions. This
is surprising in view of the recent observation that there can be no scalar
hair for a positive cosmological constant and for a large class of
potentials which, however, do not include the form of coupling considered
here \cite{Cai-Ji,Winstanley}.

The conformal coupling for the scalar field is the unique prescription that
guarantees the validity of the equivalence principle in curved spacetime 
\cite{Faraoni-Sonego}. From a different point of view, conformally coupled
scalar fields have been shown to play an important role in diverse settings
of current physical interest \cite
{Callan-Coleman-Jackiw,Peebles-Vilenkin,Seiberg-Witten}.

For a negative cosmological constant, exact black hole solutions with a
nontrivial conformally coupled scalar field, which is regular everywhere,
are known only in three dimensions \cite
{Martinez-Zanelli,Henneaux-Martinez-Troncoso-Zanelli}.

We begin with the electrically neutral case. The action is 
\begin{equation}
I[g_{\mu \nu },\phi ]=\int d^{4}x\sqrt{-g}\left[ \frac{R-2\Lambda }{16\pi G}-%
\frac{1}{2}g^{\mu \nu }\partial _{\mu }\phi \partial _{\nu }\phi -\frac{1}{12%
}\,R\,\phi ^{2}-\alpha \phi ^{4}\right] \,,  \label{Action}
\end{equation}
where $\alpha $ is a dimensionless constant ($\hslash =c=1$).

The field equations are 
\begin{equation}
G_{\mu \nu }+\Lambda g_{\mu \nu }=8\pi G\,T_{\mu \nu }\;,  \label{Eeq}
\end{equation}
\begin{equation}
\square \phi -\frac{1}{6}R\phi -4\alpha \phi ^{3}=0\;,  \label{Feq}
\end{equation}
where $\square \equiv g^{\mu \nu }\nabla _{\mu }\nabla _{\nu }$, and 
\begin{equation}
T_{\mu \nu }=\partial _{\mu }\phi \partial _{\nu }\phi -\frac{1}{2}g_{\mu
\nu }g^{\alpha \beta }\partial _{\alpha }\phi \partial _{\beta }\phi +\frac{1%
}{6}\left[ g_{\mu \nu }\square -\nabla _{\mu }\nabla _{\nu }+G_{\mu \nu
}\right] \phi ^{2}-\alpha g_{\mu \nu }\phi ^{4}\,.  \label{Tuv}
\end{equation}
Since the matter part of the action is invariant under conformal
transformations, 
\begin{equation}
g_{\mu \nu }\rightarrow \Omega ^{2}(x)g_{\mu \nu },\qquad \phi \rightarrow
\Omega ^{-1}(x)\phi ,
\end{equation}
the stress tensor is traceless, and as a consequence, the scalar curvature
is constant: 
\begin{equation}
R=4\Lambda .  \label{R}
\end{equation}

The field equations are solved by the following line element:
\begin{equation}
ds^{2}=-\left[ -\frac{\Lambda }{3}r^{2}+\left( 1-\frac{GM}{r}\right)
^{2}\right] dt^{2}+\left[ -\frac{\Lambda }{3}r^{2}+\left( 1-\frac{GM}{r}%
\right) ^{2}\right] ^{-1}dr^{2}+r^{2}d\Omega ^{2}\,,  \label{ds}
\end{equation}
where $0\leq r<\infty $, $d\Omega ^{2}$ is the metric of $S^{2}$, and the
scalar field is given by 
\begin{equation}
\phi (r)=\sqrt{\frac{3}{4\pi }}\;\frac{\sqrt{G}M}{r-GM}\,.  \label{phi}
\end{equation}
This solution exists only for $\alpha =-\frac{2}{9}\pi \Lambda G$, and
describes a static and spherically symmetric black hole, provided the
cosmological constant $\Lambda $ is positive, and the mass satisfies $%
0<GM<l/4$, where the cosmological radius is $l=\sqrt{3/\Lambda }$.

The geometry of this black hole (\ref{ds}) is the same as the
Reissner-Nordstr\"{o}m-de Sitter solution for the electric charge equal to
the mass. The inner, event, and cosmological horizons satisfy $%
0<r_{-}<GM<r_{+}<\frac{l}{2}<r_{++}<l$, where 
\begin{eqnarray}
r_{-} &=&\frac{l}{2}(-1+\sqrt{1+4GM/l})\;, \\
r_{+} &=&\frac{l}{2}(1-\sqrt{1-4GM/l})\;, \\
r_{++} &=&\frac{l}{2}(1+\sqrt{1-4GM/l})\;.
\end{eqnarray}
There is a curvature singularity at the origin, and the simple pole in the
scalar field is located at $r=GM$, which lies between $r_{-}$ and $r_{+}$.

The massless solution has a vanishing scalar field and corresponds to de
Sitter spacetime, which has a cosmological horizon at $r_{++}=l$. For the
maximum allowed value of the mass, $M=l(4G)^{-1}$, the event and
cosmological horizons coalesce, $r_{+}=r_{++}=l/2$. In case of negative mass
or for $M>l(4G)^{-1}$ the singularities become naked, thus these values of $%
M $ are excluded by cosmic censorship.

Note that the scalar field does not endow the black hole with a new hair
because $\phi $ cannot be switched off keeping the mass fixed. In fact, $M$
is the only integration constant and for $\phi \rightarrow 0$, the geometry
approaches de Sitter spacetime. Furthermore, for a given mass there are two
fairly different static and spherically symmetric black hole solutions: one
with $\phi =0$, which is the standard Schwarzschild-de Sitter black hole,
and the new one with metric (\ref{ds}), and $\phi =\phi (r)$ given by Eq. (%
\ref{phi}), which only exists for positive $\Lambda $ and a quartic
self-coupling with $\alpha =-\frac{2}{9}\pi \Lambda G$.

In the vanishing cosmological constant limit $\Lambda \rightarrow 0$, the
geometry is described by the extremal Reissner-Nordstr\"{o}m metric 
\begin{equation}
ds^{2}=-\left( 1-\frac{GM}{r}\right) ^{2}dt^{2}+\left( 1-\frac{GM}{r}\right)
^{-2}dr^{2}+r^{2}d\Omega ^{2}\,,  \label{RN}
\end{equation}
which has coalesced inner and event horizons at $r_{+}=r_{-}=GM$. However,
as the scalar field remains unchanged in this limit, it diverges at the
horizon. This asymptotically flat solution only exists for $\alpha =0$ and
was previously found independently by Bronnikov, Melnikov, and Bocharova 
\cite{Bocharova-Bronnikov-Melnikov}, and Bekenstein \cite{Bekenstein}.

$\circ $\textit{\ Electrically charged scalar black hole}

An electrically charged black hole supporting a nontrivial scalar field is
found, adding\ to the action the Maxwell term
\[
-\frac{1}{16\pi }\int d^{4}x\sqrt{-g}F^{\mu \nu }F_{\mu \nu }\;.
\]
This static and spherically symmetric solution is described by the same
metric as in Eq. (\ref{ds}), where the only nonvanishing component of the
electromagnetic field is 
\begin{equation}
F_{rt}=\partial _{r}A_{t}=\frac{Q}{r^{2}}\;,  \label{E}
\end{equation}
where the charge-to-mass ratio is given by 
\begin{equation}
\left( \frac{Q}{M}\right) ^{2}=G\left( 1+\frac{2\pi \Lambda G}{9\alpha }%
\right) \;,  \label{Q/M}
\end{equation}
and\textbf{\ }the scalar field reads 
\begin{eqnarray}
\phi (r) &=&\sqrt{\frac{3}{4\pi }}\;\frac{\sqrt{GM^{2}-Q^{2}}}{r-GM}\;,
\label{PhiQ1} \\
&=&\sqrt{-\frac{\Lambda }{6\alpha }}\;\frac{GM}{r-GM}\;.  \label{PhiQ2}
\end{eqnarray}
Note that, Eq. (\ref{Q/M}) determines the charge-to mass ratio only in terms
of the fundamental constants of the action, and implies that the bound 
\begin{equation}
\alpha <-\frac{2}{9}\pi \Lambda G  \label{The Bound}
\end{equation}
must hold in order for the electrically charged solution\ to exist. This
solution describes a regular black hole with a nonvanishing real scalar
field provided $\Lambda >0$. Moreover, the bound (\ref{The Bound}) is
saturated for the uncharged case previously discussed.

As it occurs for the uncharged case, this solution has only one independent
integration constant, and thus, the scalar field does not add a new hair. In
the vanishing cosmological constant limit, $\Lambda =0$, the solution with a
nontrivial scalar field exists only for $\alpha =0$, and in this case the
electric charge is not related to the mass. The geometry of this solution is
described by the metric (\ref{RN}), where the electric field is given by Eq.
(\ref{E}), and the scalar field by Eq. (\ref{PhiQ1}), which diverges at the
horizon. This asymptotically flat solution was previously found in Ref. \cite
{Bekenstein}.

It can be directly checked that the dominant and strong energy conditions
hold both for the charged and uncharged solutions. Indeed, the stress energy
tensor is of type I in the classification of Ref. \cite{Hawking-Ellis}. In
an orthonormal frame, 
\begin{eqnarray*}
T^{ab} &=&\frac{1}{8\pi G}\mathrm{diag}\left( \frac{G^{2}M^{2}}{r^{4}},-%
\frac{G^{2}M^{2}}{r^{4}},\frac{G^{2}M^{2}}{r^{4}},\frac{G^{2}M^{2}}{r^{4}}%
\right)  \\
&\equiv &\mathrm{diag}(\rho ,p_{1},p_{2},p_{3}).
\end{eqnarray*}
Hence, $\rho \geq 0$, and $-\rho \leq p_{i}\leq \rho $ imply the dominant
energy condition. Moreover, $\rho +p_{i}\geq 0$ $\forall i$, and $%
T_{a}^{a}\geq 0$ imply the strong energy condition.

$\circ $\textit{\ Solutions with a constant scalar field}

Since the electromagnetic stress energy tensor $T_{\mu \nu }^{E}$ is
traceless, and by virtue of Eqs. (\ref{Feq}) and (\ref{R}), solutions with a
nonvanishing constant scalar field exist for 
\begin{equation}
\phi _{0}^{2}=-\frac{\Lambda }{6\alpha }\;,  \label{PhiC}
\end{equation}
provided $\Lambda $ and $\alpha $ has opposite signs, and for any metric
satisfying Einstein's equations 
\begin{equation}
G_{\mu \nu }+\Lambda g_{\mu \nu }=8\pi \tilde{G}\,T_{\mu \nu }^{E}\;,
\label{RegEinstein}
\end{equation}
where $\tilde{G}$ is the effective Newton constant, given by 
\begin{equation}
\tilde{G}=G\left( 1+\frac{2\pi \Lambda G}{9\alpha }\right) ^{-1}.
\label{RegNewton}
\end{equation}
The same result is obtained in the presence of any other conformally
invariant matter coupling, e.g., couplings withYang-Mills fields.

An attractive gravitational interaction requires $\tilde{G}>0$, which also
gives a well defined propagator for the graviton. This implies that the
bound 
\begin{equation}
1+\frac{2\pi \Lambda G}{9\alpha }>0  \label{ExtendedBound}
\end{equation}
must hold, which in the case of a negative cosmological constant means that the $%
\alpha >-2\pi \Lambda G/9$, and for a positive cosmological constant, the
bound (\ref{The Bound}), imposed by the charge-to-mass ratio, is recovered
from a different point of view.

\acknowledgments

Useful discussions with Andr\'{e}s Gomberoff, Mokhtar Hassa\"{\i}ne and
Claudio Teitelboim are gratefully acknowledged. This work is partially
funded by FONDECYT grants 1020629, 1010446, 1010449, 1010450, 7010446, and
7010450, and from the generous support to CECS by Empresas CMPC. CECS is a
Millennium Science Institute.

\end{document}